\documentclass[aps,prl,twocolumn,groupedaddress,showpacs,superscriptaddress,amssymb,amsmath,noeprint,longbibliography]{revtex4-2}
\usepackage{graphicx}
\usepackage{dcolumn}
\usepackage{bm}
\usepackage{hyperref}
\usepackage{cleveref}
\hypersetup{
    colorlinks=true,
    linkcolor=blue,
    urlcolor=blue,
	citecolor=blue
}
\usepackage{comment}
\usepackage{color}
\usepackage[utf8]{inputenc}
\usepackage{graphicx}
\usepackage{tabularx}
\usepackage{xcolor}
\usepackage{amsmath}
\usepackage{dcolumn}
\usepackage{bm}
\usepackage{epsf}
\usepackage{braket}
\usepackage{tensor}
\usepackage{soul}
\usepackage{bbm}
\usepackage{tikz}
\usepackage[normalem]{ulem}

\newcommand\f{\frac}

\begin{document}

\title{Detection of Mpemba effect through good observables in open quantum systems}

\author{Pitambar Bagui}
\email{pitambar.bagui@students.iiserpune.ac.in} 
\affiliation{Department of Physics, Indian Institute of Science Education and Research, Pune 411008, India}

\author{Arijit Chatterjee}
\email{arijitchattopadhyay01@gmail.com} 
\affiliation{Department of Physics and NMR Research Center, Indian Institute of Science Education and Research, Pune 411008, India}

\author{Bijay Kumar Agarwalla}
\email{bijay@iiserpune.ac.in}
\affiliation{Department of Physics, Indian Institute of Science Education and Research, Pune 411008, India}

\date{\today} 

\begin{abstract}
The Mpemba effect refers to the anomalous relaxation of a quantum state that, despite being initially farther from equilibrium, relaxes faster than a closer counterpart. Detecting such a quantum  Mpemba effect typically requires full knowledge of the quantum state during its time evolution, which is an experimentally challenging task since state tomography becomes exponentially difficult as system size increases. This poses a significant obstacle in studying Mpemba effect in complex many-body systems.  In this work, we demonstrate that this limitation can be overcome by identifying suitable observables that signal rapid relaxation. Moreover, as long as the system equilibrates to a known unique steady-state, it is possible to fully detect the occurrence of quantum Mpemba effect just by measuring the observable for known state preparations. Our approach thus significantly reduces experimental complexity and offers a practical route for observing the quantum Mpemba effect in complex and extended multi-qubit setups. 
\end{abstract}

\maketitle

{\it Introduction.--} The Mpemba effect refers to an anomalous relaxation phenomenon \cite{E_B_Mpemba_1969,10.1119/1.1975687,Greaney_Jeffrey_2011,Lasanta_santos_Prl_2025,PhysRevLett.129.138002,Holtzman_roi_comm_2022,Kumar2020_nature}, wherein a system that is initially far from equilibrium relaxes much faster than a system that starts closer to equilibrium. Understanding the origin of the quantum Mpemba effect (QME) has recently attracted considerable research interest, spanning theoretical studies in closed quantum systems, particularly in the context of dynamical symmetry restoration \cite{Ares_calarese_2023_Nature,Murciano_calabrese_2024_JSM,Rylands_Bruno_2024_PRL,Lastres_calabrese_JSM_2025,Shion_calabrese_2024_PRB,Chalas_calabrese_2024_JSM,Ares_Sara_2025_PRB,Yamashika_filiberto_2025_PRA,PhysRevLett.133.140405,Hallam_clerk_2025,Bhore_clerk_2025_prbl,Ares_2025,fujimura_2025,Yu_2025}, as well as investigations in open quantum systems \cite{PhysRevLett.127.060401,PhysRevLett.131.080402,PhysRevA.110.022213,PhysRevLett.133.140404,PhysRevResearch.6.033330,PhysRevLett.133.136302,PhysRevA.110.042218,longhi2024bosonic,Longhi:24,PhysRevLett.134.220402,qj8n-k5j2,PhysRevA.111.022215,Bao_Hou_2025_PRL,PhysRevLett.134.220403,Boubakour_thomas_2025_JSM,Furtado_santos2025_annals_phys,Ali_Haddadi_2025,Israel_Marija_PRX_2019,Lu_Raz_2017_Apl,Liu_Wang_2025,Vu_Hisao_Prl_2025,teza2025,Boubakour_2025,Ivander_segal_PRE_2025,mondal_sen2025,wei_pan2025,Yan_xingli_2025_pra,summer_goold_2025} and emerging experimental realizations~\cite{Liang_2025,Zhang2025,PhysRevLett.133.010403,chatterjee2025direct,PhysRevLett.133.010402}.  To study the Mpemba effect, one typically introduces a distance function \cite{breuer2002theory,nielsen2010quantum,PhysRevLett.127.060401,PhysRevLett.133.136302,Longhi:24,PhysRevLett.133.140404}, denoted as $\mathcal{D}(\rho_t,\rho_{\mathrm{ss}})$, which quantifies the distance between the time-evolved quantum state $\rho_t$ and its steady state $\rho_{\mathrm{ss}}$.  For two distinct initial states, $\rho_0$ and $\widetilde{\rho}_0$, although initially  
$\mathcal{D}(\widetilde{\rho}_0,\rho_{\mathrm{ss}})>\mathcal{D}(\rho_0,\rho_{\mathrm{ss}})$, 
there exists a characteristic time scale $t^{\prime}$ beyond which 
$\mathcal{D}(\widetilde{\rho}_{t^{'}},\rho_{\mathrm{ss}})<\mathcal{D}(\rho_{t^{'}},\rho_{\mathrm{ss}})$. The crossing of relaxation trajectories signifies the occurrence of QME and forms the basis for its identification both theoretically, as well as experimentally.

QME has recently been observed experimentally across diverse platforms, including trapped ions \cite{PhysRevLett.133.010403,Zhang2025}, nuclear spin systems involving only single \cite{schnepper2025experimentalobservationapplicationgenuine} and two-qubit systems \cite{chatterjee2025direct}. In all such studies, the detection of QME relied on full quantum state tomography \cite{Rainer_Blatt_stricker_2022_prx,Lanyon_Blatt_nature_2017,Schmied_JMO2016} to construct an appropriate distance measure at each time instant during the evolution. However, such an approach becomes increasingly impractical as the system size grows, since the number of required measurements scale exponentially with the dimension of Hilbert space. As a result, experimental realization of QME for complex systems with many degrees of freedom -- such as in a multi-qubit setup-- poses a fundamentally challenging task \cite{chatterjee2025direct}.

In this Letter, we show how to circumvent the full quantum state tomography in order to detect quantum state Mpemba in Markovian dissipative systems. We rely on capturing this phenomena by choosing an appropriate operator that satisfy a specific condition, which first hands over the information whether quicker relaxation in state has happened or not. In addition to this information, with only the knowledge of initial and final quantum state, the occurrence of quantum state Mpemba can be fully inferred. This approach thus bypass the more expensive procedure of quantum state tomography as a function of time and thereby allowing to investigate QME in complex many-body dissipative setups. In addition, when the task of experimentally measuring the operator becomes challenging due to its non-local nature \cite{vallejo_2025}, we show how to bypass this problem and by measuring only local operators, even for extended systems, the QME can be detected.
Notably, there have been a fewer earlier attempts to investigate QME through observables \cite{ulčakar_zala_arxiv_2025,PhysRevLett.133.136302,Pemart_salas_prl_2024}. However, a clear understanding about what types of operators can carry or miss the signature of QME is completely missing. Needless to mention, our approach is independent of the choice of the distance measure, such as the trace distance \cite{PhysRevLett.133.136302,PhysRevLett.134.220403,bao2025acceleratingquantumrelaxationtemporary}, Frobenius distance \cite{PhysRevLett.127.060401}, relative entropy \cite{chatterjee2025direct,PhysRevLett.133.140404} which, are used in the context of detecting QME.

\vspace{0.1cm}
{\it Detection of QME via suitable operators.--}
In this work, we consider the evolution of the quantum states governed by Markovian dynamics, modeled through the 
Gorini - Kossakowski - Sudarshan - Lindblad (GKSL) quantum master equation \cite{breuer2002theory,Minganti_Cristiano_2018_PRA}, 
\begin{equation}
    \frac{d\rho}{dt}=-i[H,\rho]+\sum_{i}\gamma_i \big(L_i\rho L_i^{\dagger}-\f{1}{2}\{L_i^{\dagger}L_i,\rho\}\big),
    \label{lindblad_form}
\end{equation}
where $H$ is the system Hamiltonian, $L_i$ are the Lindblad jump operators and $\gamma_i$ denote the corresponding dissipation rates.  The formal solution of Eq.~\eqref{lindblad_form} is given as
$\rho_t=e^{\mathcal{L}t}\rho_0$, where $\mathcal{L}$ is the Liouvillian and $\rho_0$ is the initial state.
Using the spectral decomposition of $\mathcal{L}$, the state $\rho_t$ can be expressed as 
\begin{align}
    \rho_t=\rho_{\mathrm{ss}} + \sum_{i=1}^{d^2-1}\mathrm{Tr}(l_i\rho_0) \, e^{\lambda_it} \, r_i,
    \label{md eq:evolu_systm}
\end{align}
where $l_i$ ($r_i$) are the left (right) eigenmatrices with eigenvalue $\lambda_i$ corresponding to the $i$-th decay mode of the Liouvillian, and $d$ is the dimension of the Hilbert space. We note that $\lambda_0=0$ corresponds to the steady state $\rho_{\rm ss}$, and we adapt the ordering $|\lambda_1|\leq |\lambda_2| \leq |\lambda_3|\cdots$. A generic initial state $\rho_0$ overlaps with all the decay modes, and hence the slowest decay mode (SDM), associated with $\lambda_1$, dictates the long-time relaxation rate $\big[\tau_1=1/|\mathrm{Re}(\lambda_1)|\big]$ of the quantum state.
Given an initial state $\rho_0$ that overlaps with the SDM, one mechanism  for realizing the QME is to modify the state in such a way that its overlap with the SDM is suppressed \cite{PhysRevLett.133.140404} or eliminated \cite{PhysRevLett.127.060401,PhysRevLett.133.140404}. Let  $\widetilde{\rho_0}$ denote such a transformed state and if it satisfies the condition $\mathrm{Tr}[l_1 \widetilde{\rho}_0]=0$, then the contribution from the SDM vanishes and the late-time relaxation of $\widetilde{\rho}_0$ is governed by the next SDM, associated with the eigenvalue $\lambda_2$. This results in an exponential speed up in relaxation for $\widetilde{\rho_0}$ relative to the original state ${\rho_0}$. Furthermore, if initially  
$\mathcal{D}(\widetilde{\rho}_0,\rho_{\mathrm{ss}})>\mathcal{D}(\rho_0,\rho_{\mathrm{ss}})$, 
the relaxation trajectories display crossing during evolution, indicating the emergence of QME.

As discussed earlier, experimental detection of the QME typically requires doing full state tomography \cite{Rainer_Blatt_stricker_2022_prx,Lanyon_Blatt_nature_2017,Schmied_JMO2016} at each time points, which become increasingly impractical for large quantum systems.  This challenge, however, can be bypassed by identifying suitable observables whose dynamical expectation values can provide the required insight for detecting QME, thereby eliminating the need to construct the full state at all times during the time evolution. 
In general, the expectation value of an arbitrary Hermitian operator $\mathcal{O}$, evolving under the GKSL map, is given by
\begin{align}
    \big\langle \mathcal{O}(t)\big\rangle &= \mathrm{Tr}(\mathcal{O}\rho_t)\nonumber\\&=\mathrm{Tr}(\rho_{\mathrm{ss}}\mathcal{O})+\sum_{i=1}^{d^2-1}\mathrm{Tr}(l_i\rho_0)\, e^{\lambda_it}\, \mathrm{Tr}(r_i\mathcal{O}).
    \label{eq:Md expect_val}
\end{align}
Eq.~\eqref{eq:Md expect_val} shows that the relaxation time-scales associated with the expectation value of an observable, approaching to its steady-state value, is determined jointly by the overlaps  $\mathrm{Tr}(l_i\rho_0)$ and $\mathrm{Tr}(r_i\mathcal{O})$. As a result, for a generic initial $\rho_0$, that has finite overlap with all decay modes, the relaxation timescale of the observable is determined by the coefficient $\mathrm{Tr}(r_i\mathcal{O})$. 
Now considering that the speed up in relaxation for quantum state is achieved by engineering the modified initial state $\widetilde{\rho}_0$ such that
$\mathrm{Tr}(l_1 \widetilde{\rho}_0)=0$, then for the operator $\mathcal{O}$ if 
\begin{equation}
\mathrm{Tr}(r_1\,\mathcal{O}) \neq 0,
\label{operator-detection}
\end{equation}
then this particular observable can detect quicker state relaxation of $\widetilde{\rho}_0$ than $\rho_0$. 
We refer to such operators as \textit{good} operators, since they allow the accelerated relaxation to be inferred by comparing the dynamics corresponding to different initial states. Note that the evolution of the expectation value of a good operator only detects accelerated relaxation and does not capture 
whether or not the transformed state $\widetilde{\rho_0}$ is far from the steady-state, compared to $\rho_0$, in a distance measure. 
However, as long as the unique steady-state is known, along with the knowledge of initial states $\rho_0$ and $\widetilde{\rho_0}$, which are readily available in a definite preparation protocol, the presence or absence of QME can be unambiguously determined, bypassing the need of quantum state tomography as a function of time. To see this explicitly, consider the trace distance \cite{nielsen2010quantum,PhysRevLett.133.136302,PhysRevLett.134.220403}
\begin{equation}
\mathcal{D}_{\mathrm{tr}}(\rho_t,\rho_{\mathrm{ss}}) = 
\mathrm{Tr}\!\left[\sqrt{(\rho_t - \rho_{\mathrm{ss}})^{\dagger}(\rho_t - \rho_{\mathrm{ss}})}\right]
\label{eq: Trace-distance}
\end{equation}
which is a metric to quantify the distance between the evolving state and the steady state. The QME can be unambiguously identified along with the relaxation dynamics of \textit{good} operators,
if initially 
$\mathcal{D}_{\mathrm{tr}}(\widetilde{\rho}_0,\rho_{\mathrm{ss}})>\mathcal{D}_{\mathrm{tr}}(\rho_0,\rho_{\mathrm{ss}})$.
If instead initially $\mathcal{D}_{\mathrm{tr}}(\widetilde{\rho}_0,\rho_{\mathrm{ss}})=\mathcal{D}_{\mathrm{tr}}(\rho_0,\rho_{\mathrm{ss}})$, the phenomenon corresponds to accelerated relaxation \cite{zhou_Xing_2023_PRR}, where both states are equidistant from the steady state, but $\widetilde\rho_0$ relaxes faster than $\rho_0$. In contrast, when $\mathcal{D}_{\mathrm{tr}}(\widetilde{\rho}_0,\rho_{\mathrm{ss}})<\mathcal{D}_{\mathrm{tr}}(\rho_0,\rho_{\mathrm{ss}})$, the relaxation is conventional and no Mpemba-type effect can be observed.  Note that, although we illustrate here through the trace distance measure, any appropriate contractive distance measure \cite{longhi2024bosonic,PhysRevLett.133.140404,PhysRevLett.127.060401,chatterjee2025direct} can be employed along with an appropriately chosen operator satisfying Eq.~\eqref{operator-detection} to infer the presence of QME. It is important to mention that, in the cases of strong out-of-equilibrium driving, or engineered dissipative environments, the final steady state can be of non-trivial form and depart significantly from a simple Gibbs states. For such cases, a proper characterization of the state will be required by one full state tomography. Note that our approach is also applicable when QME occurs without completely bypassing the SDM $\big[\mathrm{Tr}(l_1 \widetilde{\rho}_0)=0\big]$, but rather by significantly reducing the overlap \cite{PhysRevLett.133.140404}. In what follows, we now address the procedure of finding good operators.
 
\vspace{0.1cm}
{\it Searching scheme for good and bad operators.--} 
In the case when the GKSL dynamics in Eq.~\eqref{lindblad_form} corresponds to a Davies map \cite{davies1979generators,PhysRevLett.133.140404,bagui2025acceleratedrelaxationmpembalikeeffect}, classification of operators into good and bad becomes significantly simpler. A key feature of the Davies map is that when the Hamiltonian is non-degenerate, the dynamics of population and coherences are completely decoupled in the energy eigenbasis. As a result, if both the initial states $\rho_0$ and $\tilde{\rho}_0$ are diagonal, i.e., confined to the population sector, the subsequent evolution is also restricted to the population sector. In such a scenario, any \textit{purely off-diagonal} Hermitian operator is a \textit{bad} observable for detecting accelerated relaxation of quantum state as its expectation value remains insensitive to the dynamics within the population subspace. In other words, in such cases Eq.~\eqref{operator-detection} is not satisfied. Instead,  Hermitian operators that are \textit{diagonal}, serve as potential candidates for inferring the presence of QME. Moreover, for a generic initial state containing coherence, and that overlaps with a complex (real) SDM of the Liouvillian,  purely off-diagonal (diagonal) operators serve as good operators for detecting QME.

Davies map, however, represents only a special type of map and does not capture the most general form of the GKSL dynamics. Depending on the structure of the Hamiltonian $H$ and the choice of the jump operators $L_i$, for instance, whether they act locally or globally, the resulting dynamics can couple the population and coherence sectors during the entire evolution. In such a scenario,  identifying the \textit{good} operators requires a more systematic approach. As the good operator satisfies the property given in Eq.~\eqref{operator-detection}, one possibility for constructing a good  operator, in such a general scenario, would be
to choose $\mathcal{O}=\big(r_1+r_1^{\dagger}\big)/2$, where recall that $r_1$ is the right eigenmatrix corresponds to the SDM. 
This is because $\mathrm{Tr}\big(r_1 \mathcal{O}\big)= \frac{1}{2} \mathrm{Tr}(r_1^2) + \frac{1}{2} \mathrm{Tr}(r_1^{\dagger} r_1) \neq 0$.
However, experimentally measuring such an operator in a complex setting, like for multi-qubit system, could be extremely challenging as, it can, in general, have weightage appearing from non-local basis operators. An alternate and more practical approach to avoid this complication, and to obtain a simple route to detect QME, would be to write $r_1$ in the operator basis as,
\begin{equation}
    r_1= \sum_{i=1}^{d^2-1} \, c_i \, \mathbb{B}_i, \quad c_i= \mathrm{Tr}(r_1 \, \mathbb{B}_i)
    \label{operator_basis}
\end{equation}
where $c_i$ is the overlap of the SDM with the $i$-th operator in the basis set. The basis operators satisfy the orthonormal property $\mathrm {Tr}\big(\mathbb{B}_i \mathbb{B}_j \big)= \delta_{ij}$ which helps to extract the coefficient $c_i$.
This decomposition procedure not only helps to identify the good operator to detect QME but also excludes the operators that are insensitive to QME and avoids any false detection of it. In other words, given an arbitrary operator $\mathcal{O}$, whether it is good or bad in the context of QME detection, one has to check the basis set ${\mathbb{B}_i}$ spanned by the operator $\mathcal{O}$. If the basis set of operator $\mathcal{O}$ is completely orthogonal to the set contained by $r_1$, it signifies that $\mathcal{O}$ is a bad operator. If this is not the case, then Eq.~\eqref{operator-detection} will be satisfied, and operator $\mathcal{O}$ will be a good operator.

In what follows, we provide three examples, where the Mpemba effect occurs at the state level, getting unambiguously detected by the \textit{good} operators time dynamics along with the knowledge of initial and final states.

\vspace{0.1cm}
{ \it Example 1: One-dimensional bosonic lattice with bulk dephasing.--}  As a first example, we consider a one-dimensional bosonic lattice composed of $L$ coupled sites and described by the tight-binding Hamiltonian
$ H =\sum_{n=1}^{L-1} J \left( a_{n}^{\dagger} a_{n+1} + \mathrm{h.c.} \right),
$ where $J$ denotes the tunnelling rate of bosons between adjacent sites, and $a_{n}^{\dagger}$ ($a_{n}$) are the bosonic creation (annihilation) operators at site $n$. The setup is further subjected to onsite bulk dephasing with identical rate $\gamma_i=\gamma$ and the corresponding $n$-th site jump operator is given as $L_n=a_n^{\dagger}a_n$. The dynamics is governed by Eq.~\eqref{lindblad_form}. This setup was recently studied in Ref.~\cite{Longhi:24} to show the existence of Mpemba effect by considering the difference of von Neumann entropy as a distance measure. Moreover, the same setup was recently experimentally realized in superconducting qubit architecture and the existence of QME was demonstrated \cite{Liang_2025}. 


We consider here the strong dephasing limit $\gamma\gg J$ and single photon sector. In this limit, the coherence in the site basis are strongly suppressed leading to the evolution confined entirely in the population sector $p_n=\rho_{n,n}(t)$. The population dynamics can be described by a Pauli master equation $\frac{d\mathbb{P}}{dt}=\mathcal{W}\,\mathbb{P}$ where $\mathbb{P}=(p_1, p_2, \cdots, p_L)^T$.
Each site population satisfies the following equation of motion \cite{Longhi:24,Lokesh_2025,Prosen_2023,Longhi_prl_2024},
(assuming open boundary conditions)
\begin{align}
    &\f{dp_n}{dt}=-2J_{\mathrm{e}} \, p_n+J_{\mathrm{e}} \, p_{n+1}+J_{\mathrm{e}}\, p_{n-1}~~(1<n<L),\nonumber \\
    &\f{dp_1}{dt}=-J_{\mathrm{e}} \, p_1+J_{\mathrm{e}}\,p_2,~\f{dp_L}{dt}=-J_{\mathrm{e}}\,p_L+J_{\mathrm{e}}\,p_{L-1},
    \label{pauli_rate_eqn}
\end{align}
where $J_{\mathrm{e}}=2J^2/\gamma$ is the effective incoherent hopping rate in the strong dephasing limit. As the $\mathcal{W}$ matrix is Hermitian, the steady-state corresponds to equal probability of finding the photon at each lattice site $\rho_{\mathrm{ss}}=(1/L)\sum_{i=1}^{L}\ket{i}\bra{i}$.
We analyze two distinct initial conditions, as discussed in Ref.~\cite{Longhi:24} and further realized experimentally in Ref.~\cite{Liang_2025}, to observe the QME. We consider a localized pure state $\widetilde{\rho}_0 = |n_{\rm{mid}}\rangle\langle n_{\rm{mid}}|$, where the photon is initially placed at the middle site of the lattice $n_{\rm mid} = (L+1)/2$ ($L$ is odd), and a delocalized mixed state ${\rho}_0 = \frac{1}{L-1}\sum_{i=1}^{L-1}|i\rangle\langle i|$ which has equal density at all sites except the last site. It is therefore intuitively clear that the state ${\rho}_0$ is closer to the steady-state than the state $\widetilde{\rho}_0$. As $\mathcal{W}$ is tridiagonal, it can be easily diagonalized and the population at the $n$-th site at time $t$ can be obtained as, $p_n(t)=\frac{1}{L}+\sum_{\alpha=2}^L \Delta_{\alpha}u_n^{(\alpha)}\exp(-|\lambda_{\alpha}|t)$, where $u_n^{(\alpha)}=\sqrt{2/L}\cos{\big[(\alpha-1)(2n-1)\pi/2L\big]}$, $2 \leq \alpha \leq L$, is the amplitude of the $\alpha$-th decay mode at the site $n$ and $\Delta_{\alpha}=\sum_{n=1}^{L}u_n^{(\alpha)}p_n(0)/\sum_{n=1}^{L}\big[u^{(\alpha)}_n\big]^2$ is the projection amplitude of the initial state $p_n(0)$ on the $\alpha$-th decay mode of $\mathcal{W}$. Interestingly, the SDM which corresponds to $u_n^{(2)}$, possesses a node at the middle site of the lattice. 
As a consequence, the localized pure state $\widetilde{\rho}(0)$ does not couple or excite the mode $u_n^{(2)}$. In contrast, the mixed state ${\rho}(0)$ has a finite overlap with the SDM, resulting in a slower relaxation toward the stationary state. Thus, the farther state $\widetilde{\rho}(0)$ relaxes faster than $\rho(0)$. Following our prescription of detecting QME using operators, 
for the given setup, dynamics of any local density operator $\mathcal{O}=|i\rangle \langle i|$ that measures the local site density $\langle n_i \rangle =p_i(t)$ can detect the accelerated relaxation,  except the middle site population $(i \neq \rm{mid})$. For the middle site case with operator $\mathcal{O}=|n_{\rm mid}\rangle \langle n_{\rm mid}|$, following Eq.~\eqref{operator-detection}, ${\rm Tr} \big(r_1 \mathcal{O}\big)= u_{\rm mid}^{(2)} = \sqrt{2/L}\cos{[\pi/2]}=0$. This is precisely due to the presence of a node of the SDM $u_{n}^{(2)}$, at the middle site of the lattice.

\begin{figure}
    \centering
\includegraphics[width=\columnwidth]
{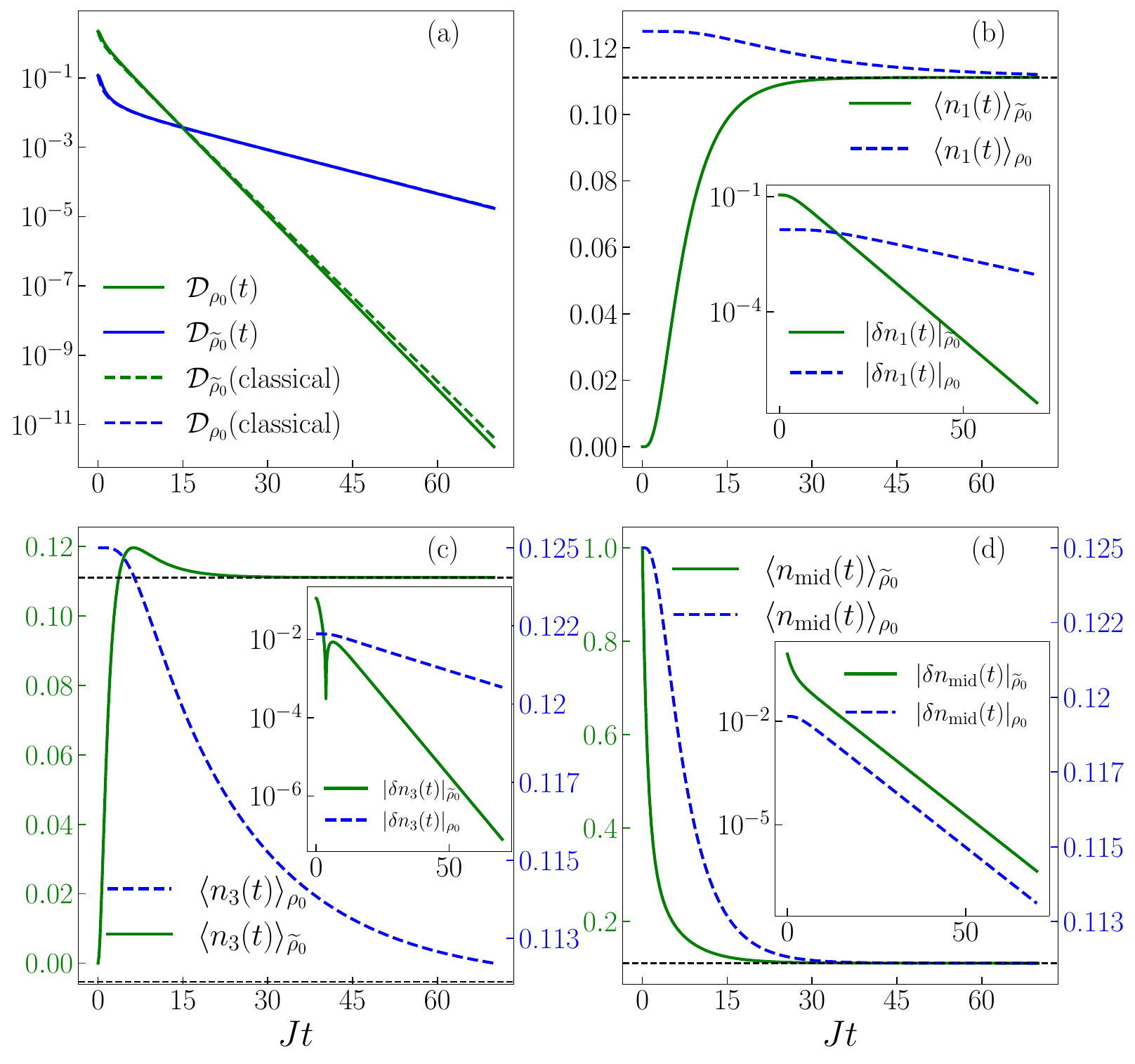}
    \caption{Plot for the detection of the QME through observables for a one-dimensional lattice subjected to bulk dephasing. (a) QME captured through the distance measure $\mathcal{D}_{\rho_t}(t)=S(\rho_{\mathrm{ss}})-S(\rho_t)$, defined as the von Neumann entropy difference between the steady state and the time-evolved state, following exact GKSL dynamics and the dynamics using effective equation involving populations as given in Eq.~\eqref{pauli_rate_eqn}. Here ${\rho}_0 = \frac{1}{L-1}\sum_{i=1}^{L-1}|i\rangle\langle i|$ represents a delocalized initial state and $\widetilde{\rho}_0 = |n_{\rm{mid}}\rangle\langle n_{\rm{mid}}|$, 
    represents a localized initial state in the middle site of the lattice. Even though $\widetilde{\rho}_0$ is far from the steady-state it relaxes faster compared to delocalized state $\rho_0$. 
    (b) and (c): Inferring the existence of QME by tracking the dynamics of \textit{good} operators, which in this case are local density (populations) for any site except the middle site. The expectation values of $n_1=a_1^{\dagger}a_1$, $n_3=a_3^{\dagger}a_3$, and their absolute deviation from the steady state $|\delta n_1 (t)|$ (inset),$|\delta n_3 (t)|$ (inset) both capture this quicker acceleration. (d) The expectation value of $n_5=a_5^{\dagger}a_5$ and $|\delta n_5|$ does not capture the quick acceleration due to the presence of a node for the SDM at the middle site of the lattice. Here we choose the parameters as $L=9$, $J=1$, and $\gamma=5$.}
\label{lattice_with_dephasing_Mpemba_detect}
\end{figure}

In Fig.~\ref{lattice_with_dephasing_Mpemba_detect}, we show the results for the detection of the QME through observables for the lattice setup. In Fig.~\ref{lattice_with_dephasing_Mpemba_detect} (a), we plot the distance measure $
\mathcal{D}_{\rho_t}(t)=S(\rho_{\mathrm{ss}})-S(\rho_t)$, defined as the von Neumann entropy \big[$S(\rho)= -{\rm Tr} \big[\rho \ln \rho \big]$\big] difference between the steady state and the time-evolved state, following exact dynamics (solid line) and the dynamics using effective equation (dashed line) involving populations, as given in Eq.~\eqref{pauli_rate_eqn}.
This measure was employed in Ref.~\cite{Longhi:24,Liang_2025} to detect the Mpemba effect for this setup.
In Fig.~\ref{lattice_with_dephasing_Mpemba_detect}(b) and Fig.~\ref{lattice_with_dephasing_Mpemba_detect}(c), we show that tracking the local density evolution of any lattice sites other than the middle site can detect the accelerated relaxation between $\rho_0$ and $\widetilde{\rho_0}$. We plot in the insets the late time dynamics of the observables $|\delta n_i(t)|= |\langle n_i(t)\rangle - \langle n_i\rangle_{\rm ss}|$, clearly demonstrating the accelerated relaxation.
In Fig.~\ref{lattice_with_dephasing_Mpemba_detect}(d), we show the local site population of the middle site does not qualify as a good operator for detecting QME. This is precisely because of the reflection symmetry of the lattice about the origin which leads to a node for the SDM at the middle site of the lattice. The inset further confirms the same asymptotic relaxation time scale. In other words, for this lattice setup, the measurement of the local site population, except the middle site, along with the knowledge of state preparation, can unambiguously detect the occurrence of QME, which otherwise would require a full state tomography in time.

\begin{figure}
    \centering
\includegraphics[width=\columnwidth]{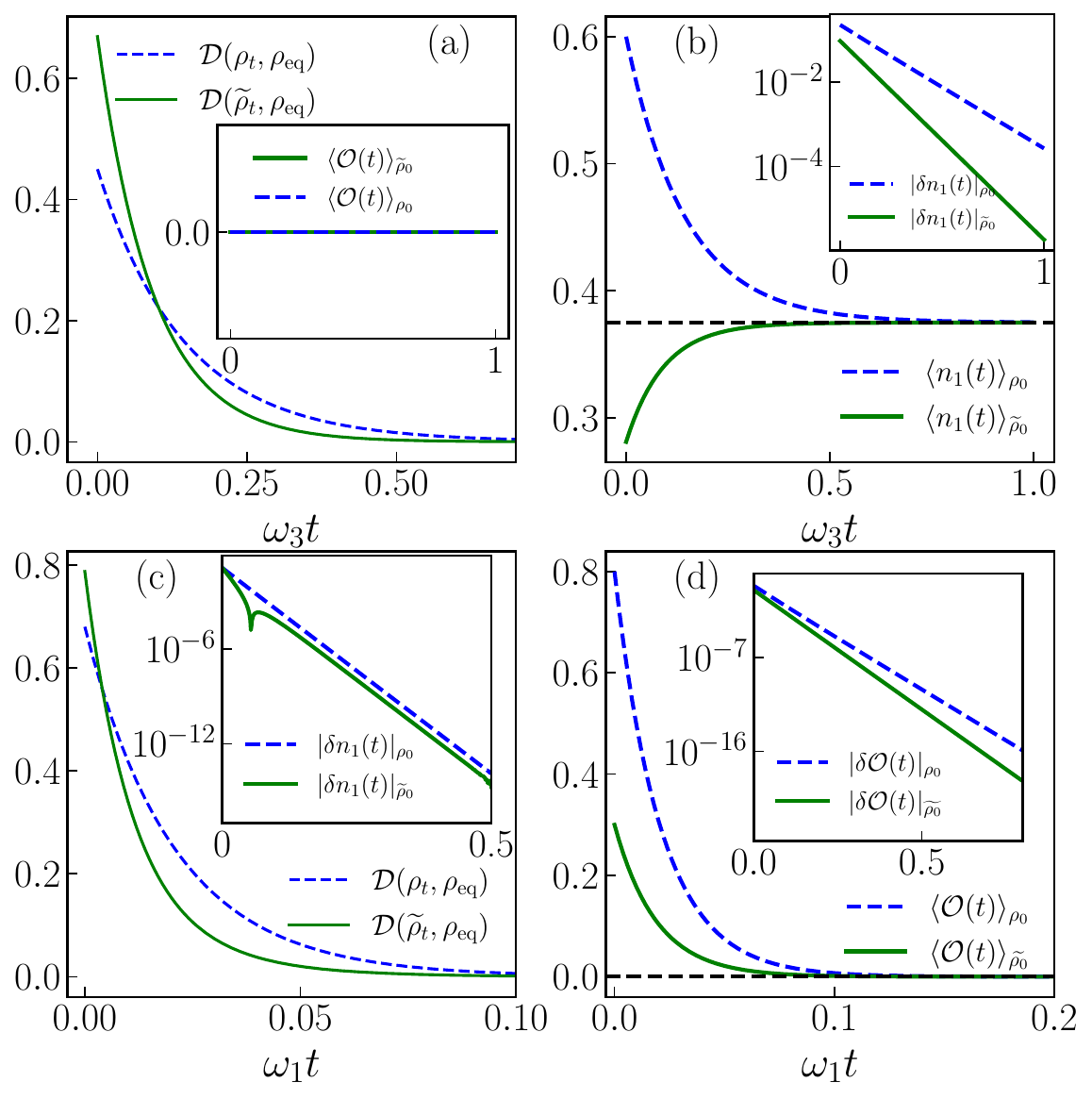}
    \caption{Plot for detection of the QME through observable for an $N$-level system evolving under the Davies map. For numerics, we consider here $N=3$. For (a) and (b), we choose diagonal initial states for both  $\rho_0$ and $\widetilde{\rho}_0$. $\rho_0$ overlaps with the SDM of the population sector, whereas $\widetilde{\rho}_0$ does not. In addition, $\widetilde{\rho_0}$ is far from the thermal state $\rho_{\mathrm{ss}} = {e^{-\beta H}}/{\mathcal{Z}}$. As a result, QME emerges and crossing in trace distance measure [Eq.~\eqref{eq: Trace-distance}] is observed. To detect this effect through operators, the inset of (a) shows that purely off-diagonal operators, such as  
    $\mathcal{O}_{ij}=1$ for $i \neq j$, do not carry any accelerated relaxation signature as they always remain orthogonal to the population subspace. Whereas (b) shows that the diagonal operators, such as the population of the ground state $\langle n_1(t)\rangle$, shows a clear acceleration for the transformed state $\widetilde{\rho_0}$ (see inset for late time relaxation in a semi-log scale). For (c) and (d), we choose initial states with coherences for both $\rho_0$ and $\widetilde{\rho}_0$. In this case, $\rho_0$ overlaps with the SDM of the Liouvillian, given as complex conjugate pairs, whereas $\widetilde{\rho}_0$ does not. In this case as the insets show, diagonal operators, such as $\langle n_1(t)\rangle$ do not detect accelerated relaxation, whereas purely off-diagonal operators, such as $\mathcal{O}$, defined above, can clearly detect the acceleration. We choose the parameters $\omega_{1 (2) (3)}=0.3 \, (0.7) \, (1)$, $\gamma=0.4$, $T=3$ for plots (a) and (b) and for plots (c) and (d) we choose $\omega_{1 (2) (3)}=1 \, (1.1)\, (1.2)$, $\gamma=1$, and $T=3$.}
    \label{qtrit_Mpemba_detect_diag_rho}
\end{figure}

\vspace{0.1cm}
{ \it Example 2: $N$-level system.--} As a second example, we consider a $N$-level quantum system described by the Hamiltonian
$ H = \sum_{i=1}^{N} \omega_i \ket{i}\bra{i},
$
where $\ket{i}$ is an energy eigenstates of $H$ with eigenvalue $\omega_i$.  The system interacts weakly with a thermal reservoir, leading to incoherent transitions between different energy levels. 
The corresponding jump operators are given by
$L^{-}_{mn} = \sqrt{\gamma \,  \big(1 + n_b (\omega_{mn})\big)}\, \ket{n}\bra{m}, 
~L^{+}_{mn} = \sqrt{\gamma \, n_b (\omega_{mn})}\, \ket{m}\bra{n},
$
where $\ket{m}$ and $\ket{n}$ denote the eigenstates connected by the bath induced transitions, and 
$n_b(\omega_{mn}) = [\exp(\beta \omega_{mn}) - 1]^{-1}$ is the Bose-Einstein occupation number of the bath mode at frequency 
$\omega_{mn} = \omega_m - \omega_n$ with corresponding inverse bath temperature $\beta=1/k_bT$. The jump operators satisfy the quantum detailed balance condition, ensuring that the dynamics corresponds to a Davies map~\cite{davies1979generators,PhysRevLett.133.140404,bagui2025acceleratedrelaxationmpembalikeeffect}. Consequently, the steady state of the evolution is the thermal (Gibbs) state
$\rho_{\mathrm{ss}} = {e^{-\beta H}}/{\mathcal{Z}}$, with $\mathcal{Z} = \mathrm{Tr}(e^{-\beta H})$ being the partition function. 
Moreover, when the Hamiltonian is non-degenerate, which is the case here, the Davies map takes a block-diagonal form in the energy eigenbasis: one block governs the relaxation of populations towards the Gibbs state. The other block on the other hand, captures the exponential decay of coherence elements. Moreover, The right (left) eigenmatrices $r_i$ ($l_i$) corresponding to the population sub-block are purely diagonal, whereas those associated with the coherence sub-block are purely off-diagonal. 

In Fig.~\ref{qtrit_Mpemba_detect_diag_rho} (a) and (b), we consider two diagonal initial states $\rho_0$, and $\widetilde\rho_0$, for a three level $(N=3)$ system. The state $\rho_0$ overlaps with the SDM, present in the population sector, whereas $\widetilde{\rho_0}$ does not. In (a) the crossing in the dynamics of trace distance measure [Eq.~\eqref{eq: Trace-distance}] confirms the existence of QME. To infer about this through operators, the inset shows the dynamics of a purely off-diagonal operator $\mathcal{O}$ with $\mathcal{O}_{ij}=1$ for $i\neq j$, $i, j=1,2, 3$ which does not capture any signature of accelerated relaxation. In contrast, diagonal operators, such as population of the ground state $\langle n_1 (t) \rangle$ clearly shows rapid relaxation, as plotted in (b). In (c) and (d), we consider the alternate scenario, where the initial states $\rho_0$ and $\widetilde{\rho}_0$ contain coherences in the energy eigebasis. In addition, $\rho_0$ overlaps with the SDM of the Liouvillian, given as complex conjugate pairs, whereas $\widetilde{\rho}_0$ does not overlap with the SDM. In this case, as shown in the insets of both the plots, diagonal operators, such as the one that measures ground state population $\langle n_1(t)\rangle$ do not detect accelerated relaxation, whereas purely off-diagonal operators, such as $\mathcal{O}$ defined before as $\mathcal{O}_{ij}=1$ for $i\neq j$, can clearly detect the acceleration. This confirms our analysis that for the Davies map, depending on the initial states and SDM of Liouvillian, the good operators could either be the diagonal or purely off-diagonal.

\begin{figure}
    \centering
    \includegraphics[width=\columnwidth]{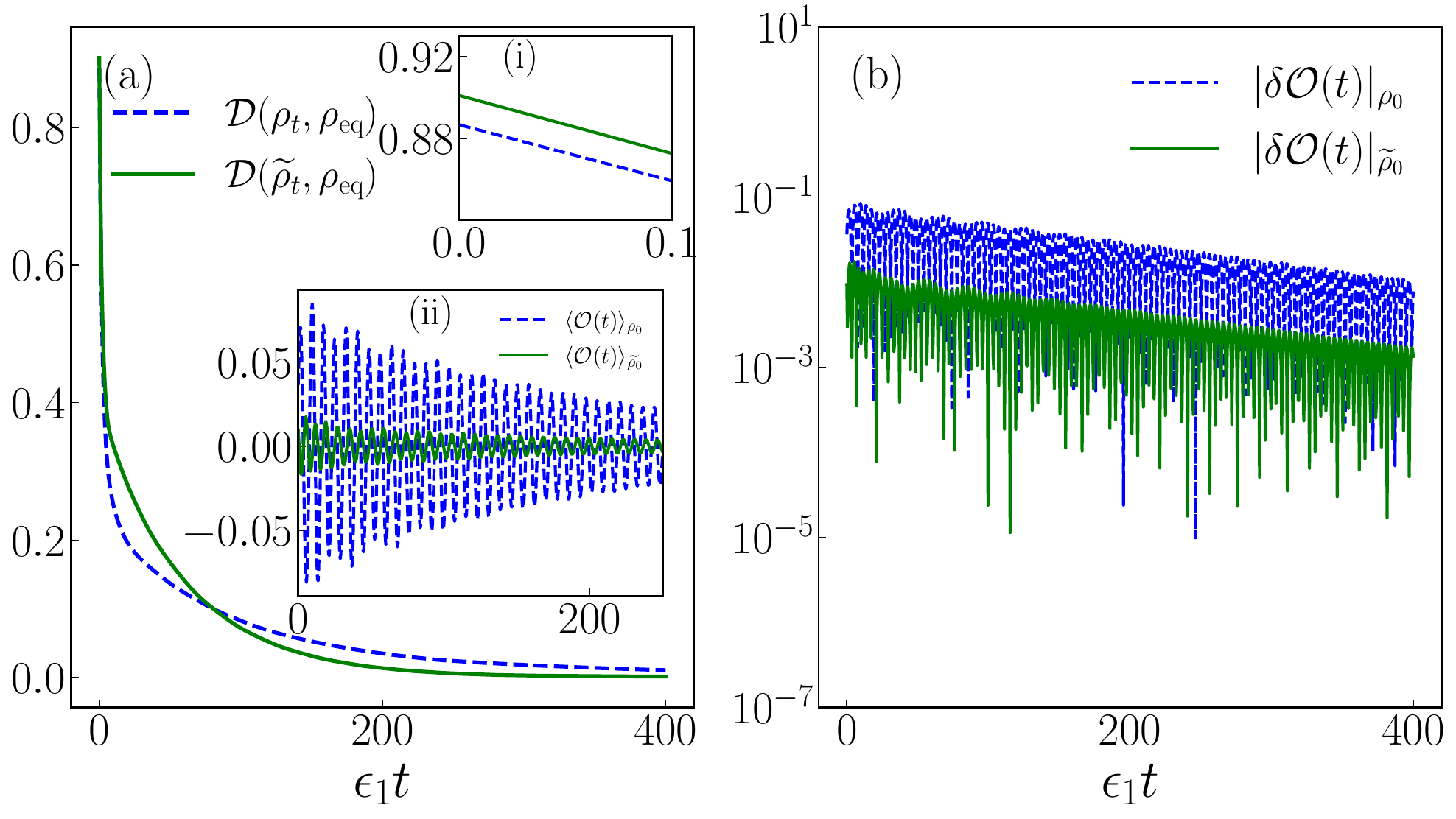}
    \caption{Plot for detection of the QME in a boundary-driven $N$-qubit setup with $N=4$. The QME occurring at the quantum state level, as shown in (a) by plotting the trace distance [Eq.~\eqref{eq: Trace-distance}] (inset (i) shows the zoomed version at early times) is now being detected through tracking the dynamics of a local operator $\mathcal{O}=\mathbb{I}\otimes\sigma_x\otimes \mathbb{I}\otimes\mathbb{I}$, as shown in the inset (ii). For the numerics, we choose a random state $\rho_0$ as the initial state and obtain the transformed state $\widetilde{\rho}_0$ by applying a unitary operation using the Metropolis protocol, as discussed in Ref.~\cite{PhysRevLett.133.140404}. The transformed state $\widetilde{\rho}_0$ has a higher distance from the steady state in the trace distance measure [see inset (i)], but as the overlap with the SDM is reduced via the unitary protocol, QME emerges. In (b) we plot the dynamics of the
    absolute deviation of the operator from the steady state $|\delta \mathcal{O}(t)|$ and a clear accelerated relaxation for the case of transformed initial state $\widetilde{\rho_0}$ is observed. We choose the parameters $g=0.05, \gamma_1=0.2,\gamma_N=0.5,~\epsilon_{1\,(2)\,(3)\,(4)}=1\,(0.8)\,(0.6)\,(0.5),~T_1=5,T_N=2,\mu_1=\mu_N=0$ for the simulation.
    \label{multi-qubit}
}
\end{figure}

\vspace{0.1cm}
{\it Example 3: Boundary driven multi-qubit system:--}
As a next example, we consider a boundary driven multi-qubit setup in which a system with $N$ qubits, described by the Hamiltonian,  $H=\sum_{i=1}^{N}\epsilon_i \, \sigma_{+}^{(i)}\sigma_{-}^{(i)}
+ g\sum_{i=1}^{N-1} 
\big( \sigma_{+}^{(i)}\sigma_{-}^{(i+1)}
+ \sigma_{-}^{(i)}\sigma_{+}^{(i+1)} \big)$, is driven out-of-equilibrium via local jump operators at the two ends of the setup.
Here, $\epsilon_i$ is the on-site energy and $g$ describes the interaction strength between the nearest neighbours.
The dynamics of the setup is governed by Eq.~\eqref{lindblad_form} with the jump operators associated with the left end ($1$-st site) are $L_{1}^{+ (-)}=\sigma_{{+}(-)}^{(1)}$ with corresponding rates $\gamma_{1} f_{1}(\epsilon_1)$ and $\big(\gamma_{1}\left[1-f_{1}(\epsilon_1)\right]\big)$ and for the right end ($N$-th site) the jump operators are $L_{N}^{+ (-)}=\sigma_{{+}(-)}^{(N)}$, with rates $\gamma_{N} f_{N} (\epsilon_{N})$ and $\big(\gamma_{N}\left[1-f_{N}(\epsilon_{N})\right]\big)$. Here $f_i(\epsilon_i)$ is the Fermi distribution evaluated at the onsite energy $\epsilon_j$, chemical potential $\mu_j$, and the temperature $T_j$, which are generally different at the two ends and as a result drives the system out-of-equilibrium. 
 For such a boundary driven non-equilibrium setup, the population and coherences are, in general, not decoupled unlike the previous cases, such as the Davies map [example 2]. Therefore, in this scenario, to find a \textit{good} operator to detect QME, we rely on decomposing the SDM in the operator basis, as given in Eq.~\eqref{operator_basis}. 
 Interestingly, such a decomposition provides a way to look for local operators which have non-zero overlap with SDM [i.e., finite $c_i$ in Eq.~\eqref{operator_basis}] to infer about the QME. As a result, 
even in such an extended multi-qubit setup, a local observable dynamics could potentially help in detecting QME unambiguously.  



In Fig.~\ref{multi-qubit} we show the results for this setup considering $N=4$. We generate the initial state $\rho_0$ randomly and perform a unitary transformation to $\rho_0$ to reduce the overlap of several consecutive slowest decay modes $\{r_1,r_1^{\dagger},r_2,r_2^{\dagger},r_3\}$ of the Liouvillian and obtain $\widetilde{\rho_0}$. Such reduction in the overlap with the SDMs can be carried out by implementing a Metropolis algorithm, as recently adapted in Ref.~\cite{PhysRevLett.133.140404} for a similar setup. In Fig.~\ref{multi-qubit} (a), we show the crossing in the trace-distance [Eq.~\eqref{eq: Trace-distance}] at a certain time which confirms the occurrence of QME. The inset (ii) shows the rapid relaxation of a single site local basis operator $\mathcal{O}=\mathbb{I}\otimes\sigma_x\otimes \mathbb{I}\otimes\mathbb{I}$ for $\widetilde{\rho_0}$ compared to $\rho_0$ and thus helps to detect the QME. In this case, there are several local basis operators that overlap with the SDM $r_1$. However, we choose the one that has the maximum overlap. In Fig.~\ref{multi-qubit} (b), we plot the late time relaxation dynamics of this local operator $\mathcal{O}$ which clearly capture the acceleration that happens for $\widetilde{\rho_0}$ compared to $\rho_0$.
Thus, our analysis explicitly demonstrate  that for extended multi-qubit setups, even local operators can detect QME. As a result, our approach not only bypass the complex tomography procedure but also helps to bypass measuring complex non-local operators in the context of QME detection.
 
\vspace{0.1cm}
{\it Summary.--} QME is an intriguing phenomenon that has gathered significant attention in recent times. Experimental detection of this effect requires state tomography which is notoriously difficult for many-body systems. In this work, we bypass this scenario and provide a way to infer about it for Markovian systems through suitable observables. We further provide a recipe for finding such observables. We demonstrate through different examples that even for extended lattice systems (example 1 and example 3) local operators can help in detecting QME. Our study therefore not only bypass the complex tomography procedure in real time but also gives a way to bypass measuring non-local operators.

In future, it will be interesting to understand the nature of local operators in detecting QME for integrable and non-integrable systems where symmetries play a paramount role \cite{ulčakar_zala_arxiv_2025}.

\vspace{0.1cm}
\emph{Acknowledgement.--} PB acknowledges Triyasa Kunti, Aritra Saha for useful discussions and the University  Grants Commission (UGC), Government of India, for the research fellowship (Ref No.- 231620073714). AC acknowledges T.S. Mahesh for valuable discussions and Patankar Fellowship for financial support. BKA acknowledges CRG Grant No. CRG/2023/003377 from Science and Engineering Research Board (SERB), Government of India.  BKA acknowledges the hospitality of the International Centre of Theoretical Sciences (ICTS), Bangalore, India and International Centre of Theoretical Physics (ICTP), Italy, under the associateship program.

\bibliography{bibliography}

\end{document}